\documentclass[ reprint, superscriptaddress, amsmath, amssymb, aps, pre, floatfix, showkeys]{revtex4-1}

\usepackage{graphicx}
\usepackage{dcolumn}
\usepackage{bm}
\usepackage{hyperref}
\usepackage{color}

\begin{document}


\title{Ion friction at small values of the Coulomb logarithm}

\author{Tucker Sprenkle}
\author{Adam Dodson}
\author{Quinton McKnight}
\author{Ross Spencer}
\author{Scott Bergeson}
 \email{scott.bergeson@byu.edu}
\affiliation{
 Department of Physics and Astronomy, Brigham Young University, Provo, UT 84602, USA
}

\author{Abdourahmane Diaw}
\email{diaw@lanl.gov}
\affiliation{
 Computational Physics and Methods Group, Los Alamos National Laboratory, Los Alamos, New Mexico 87544, USA
}

\author{Michael S. Murillo}
\email{murillom@msu.edu}
\affiliation{
 Department of Computational Mathematics, Science and Engineering, Michigan State University, East Lansing, Michigan 48824, USA
}

\date{\today}

\begin{abstract}
Transport properties of high-energy-density plasmas are influenced by the ion collision rate. Traditionally, this rate involves the Coulomb logarithm, $\ln\Lambda$. Typical values of $\ln\Lambda$ are $\approx 10~\mbox{to}~20$ in kinetic theories where transport properties are  dominated by weak-scattering events caused by long-range forces. The validity of these theories breaks down for strongly-coupled plasmas, when $\ln\Lambda$ is of order one. We present measurements and simulations of collision data in strongly-coupled plasmas when $\ln\Lambda$ is small. Experiments are carried out in the first dual-species ultracold neutral plasma (UNP), using Ca$^+$ and Yb$^+$ ions. We find strong collisional coupling between the different ion species in the bulk of the plasma. We simulate the plasma using a two-species fluid code that includes Coulomb logarithms derived from either a screened Coulomb potential or a the potential of mean force. We find generally good agreement between the experimental measurements and the simulations. With some improvements, the mixed Ca$^+$ and Yb$^+$ dual-species UNP will be a promising platform for testing theoretical expressions for $\ln\Lambda$ and collision cross-sections from kinetic theories through measurements of energy relaxation, stopping power, two-stream instabilities, and the evolution of sculpted distribution functions in an idealized environment in which the initial temperatures, densities, and charge states are accurately known.
\end{abstract}


\doi{10.1103/PhysRevE.99.053206}
\maketitle

\section{Introduction}

Understanding energy transport and relaxation processes is an important aspect of optimizing plasma fusion experiments and determining their equations of state \cite{PhysRevLett.112.145004,0953-4075-50-13-134002,PhysRevE.93.043203}. In some systems, the electron and ion temperatures can be different by an order of magnitude or more \cite{Lindl2014}. As fusion proceeds, this energy difference is exacerbated when fusion products asymmetrically deposit their energy into the electron system due to the large mass ratio \cite{Vorberger2014}. The resulting two-temperature problem is a long-standing plasma physics issue in many systems \cite{PhysRevLett.120.204801, PhysRevE.77.026401, Cooley2016,Dharuman2017, PhysRevE.96.063310,PhysRevE.95.043202, PhysRevE.90.013104, Mabey2017, Shaffer2017, alpha1, alpha2}.

Kinetic calculations, based on the Boltzmann equation or one of its simplifications, rely on a statistical assumption about Coulomb collisions between charged particles. They are most accurate when the collisions are frequent and weak, corresponding to large impact parameters and small-angle scattering. Mathematical expressions for two-body processes such as the electron-ion collision rate or the ion-ion collision cross-section are modified by the Coulomb logarithm to account for the many-body, long-range nature of Coulomb collisions. In the Landau-Spitzer treatment, the Coulomb logarithm is conveniently written as $\ln\Lambda = \ln ( \lambda_D / r_0^{ } )$, where the electron Debye length is $\lambda_D = [\epsilon_0^{ } k_{B}^{ } T / (n  e^2)]^{1/2}$ and the classic distance of closest approach is $r_0^{ } = e^2/ (4\pi \epsilon_0^{ } k_{\rm B} T)$. The Coulomb logarithm is a function of density and temperature and it multiplies every cross-section and collision rate in kinetic calculations \cite{Huba2013}. Its value ranges from 10 to 20 for weakly coupled plasmas, and represents an average over many long-range binary collisions in the plasma.

Recent theoretical and computational work extend the Landau-Spitzer treatment described above to higher density plasmas. These treatments use the screened Coulomb interactions in which the many-body physics lacking in binary models is accounted for through the choice of an ad-hoc effective screening length
\cite{doi:10.1063/1.1724389, 1984ApJ...278..769M, 1986ApJS...61..177P, PhysRevE.93.043203}. Recently, the effective potential theory which models many-body correlation effects by treating binary interactions as arising through the potential of mean force rather than the screened Coulomb potential has been proposed in Ref. \cite{doi:10.1063/1.4967627}. All these models have been used to calculate transport properties for plasmas across coupling regimes \cite{PhysRevE.91.033104, PhysRevE.93.043203,PhysRevLett.101.135001, PhysRevLett.108.225004, PhysRevLett.110.235001, doi:10.1002/ctpp.201700109}. These are typically compared to results from molecular dynamics simulations, and, where possible, experimental data \cite{PhysRevLett.117.155001, PhysRevX.6.021021, PhysRevE.89.043107, PhysRevLett.109.185008, doi:10.1063/1.5016194}.

Ultracold neutral plasmas (UNPs) span the phase space region where the Coulomb logarithm values are small. These systems have enabled studies of plasma dynamics and evolution in a highly idealized environment \cite{KILLIAN200777,0034-4885-80-1-017001}, serving, in a way, as high-energy-density plasma  simulators \cite{Murillo2004, PhysRevLett.96.165001}. They are generated by resonantly photo-ionizing mK-temperature atoms \cite{PhysRevLett.83.4776, Rolston2003, PhysRevLett.92.143001, PhysRevLett.95.235001, Murphy2014} or molecules \cite{PhysRevLett.101.205005, PhysRevA.96.023613}. The initial plasma density ($10^7$ to $10^{12}~\mbox{cm}^{-3}$) and electron temperature (5 to 500 K) are selected with small uncertainties. Simulations \cite{PhysRevLett.88.055002, PhysRevA.70.033416}, laser spectroscopy \cite{Cummings2005, PhysRevLett.101.073202,PhysRevLett.109.185008}, radio-frequency measurements\cite{PhysRevLett.92.253003, PhysRevA.87.013410}, charged particle detection and imaging \cite{PhysRevLett.101.195002,PhysRevLett.85.318} are all used to characterize these systems. They have deepened our understanding of the time-evolving density \cite{PhysRevE.67.026414,PhysRevLett.108.065003}, electron and ion temperatures \cite{PhysRevA.87.013410,PhysRevE.87.033101,PhysRevLett.94.205003, Kuzmin2002, McQuillen2015, PhysRevE.93.023201}, collision and recombination rates \cite{PhysRevLett.99.145001,PhysRevE.91.033101}, expansion, velocity relaxation \cite{PhysRevLett.109.185008}, localization \cite{PhysRevLett.120.110601}, and self-diffusion \cite{PhysRevX.6.021021} in strongly-coupled Coulomb systems. The recent realization of laser-cooling ions in an ultracold neutral plasma open the possibility of extending all of these studies farther into the strongly coupled plasma regime \cite{Langin2019}.


In this article we present the first dual-species UNP, using Ca$^+$ and Yb$^+$ ions. We report simulations and measurements of the Ca$^+$ ion velocity distribution in the dual-species UNP. We also present a two-fluid model with two representations of the friction force between the ions. Those simulations reproduce the main features of the measured ion velocity distribution. This system provides a unique platform for future studies of collision physics in strongly-coupled plasmas. In this system it should be possible to study idealized versions of classic plasma problems such as inter-species diffusion \cite{PhysRevLett.105.115005}, multi-species plasma expansion \cite{PhysRevLett.120.074801}, two-stream instabilities, the sensitivity of bump-on-tail evolution to electron screening \cite{Aslanyan2017}, shock evolution \cite{PhysRevLett.118.025001, PhysRevLett.119.195001}, and evaluations of the Coulomb logarithm when the plasma approaches the non-ideal state \cite{PhysRevLett.110.235001, PhysRevE.93.043203}.

\section{Experimental description}

We simultaneously trap $10^7$ neutral $^{40}$Ca and $^{174}$Yb atoms in a magneto-optical trap (MOT) \cite{PhysRevLett.59.2631} at a temperature of a few mK \cite{PhysRevLett.95.235001}. The spatially-overlapped MOTs operate on the strong resonance transitions at 423 and 399 nm for Ca and Yb, respectively. Unlike dual-species MOTs with alkali atoms or combinations of alkali and alkaline-earth atoms, the $^{40}$Ca and $^{174}$Yb atoms occupy the same physical space in the MOT without adversely influencing the number of trapped atoms of either species because there is no ground-state hyperfine structure \cite{PhysRevA.73.023406, PhysRevA.84.011603, Extavour2006, Ridinger2011}. The spatial density profile is Gaussian, $n = n_0 \exp(-r^2 / 2\sigma^2)$. In order to minimize spatial inhomogeneities in the neutral atom clouds stemming from imperfect laser beams, the neutral atoms are allowed to expand for $100~\mu\mbox{s}$ before formation of the plasma.

The neutral atoms in the MOT are resonantly-ionized using ns-duration laser pulses in a two-step process. The initial electron temperature in the plasma is determined by varying the wavelengths of the 390 nm (Ca) and 395 nm (Yb) laser pulses. In experiments reported here, the electron temperature is $T_e = 96~\mbox{K}$. The ion densities are determined by varying the intensity of these same laser pulses. With our few-mJ pulses, we can ionize all of the Ca atoms and up to 60\% of the Yb atoms. The peak density of the Ca$^+$ plasma is $n_0 = 1.8\times 10^{10}~\mbox{cm}^3$ with an initial rms size of $\sigma_0 = 0.29~\mbox{mm}$. The peak density of the Yb$^+$ plasma in the experiments reported here varies from $n_0 = 0.2 \times 10^{10}~\mbox{cm}^{-3}$ to $n_0 = 1.8 \times 10^{10}~\mbox{cm}^{-3}$ with an initial rms size of $\sigma_0 = 0.37~\mbox{mm}$.

\begin{figure}
  \centerline{\includegraphics[width=0.95\linewidth]{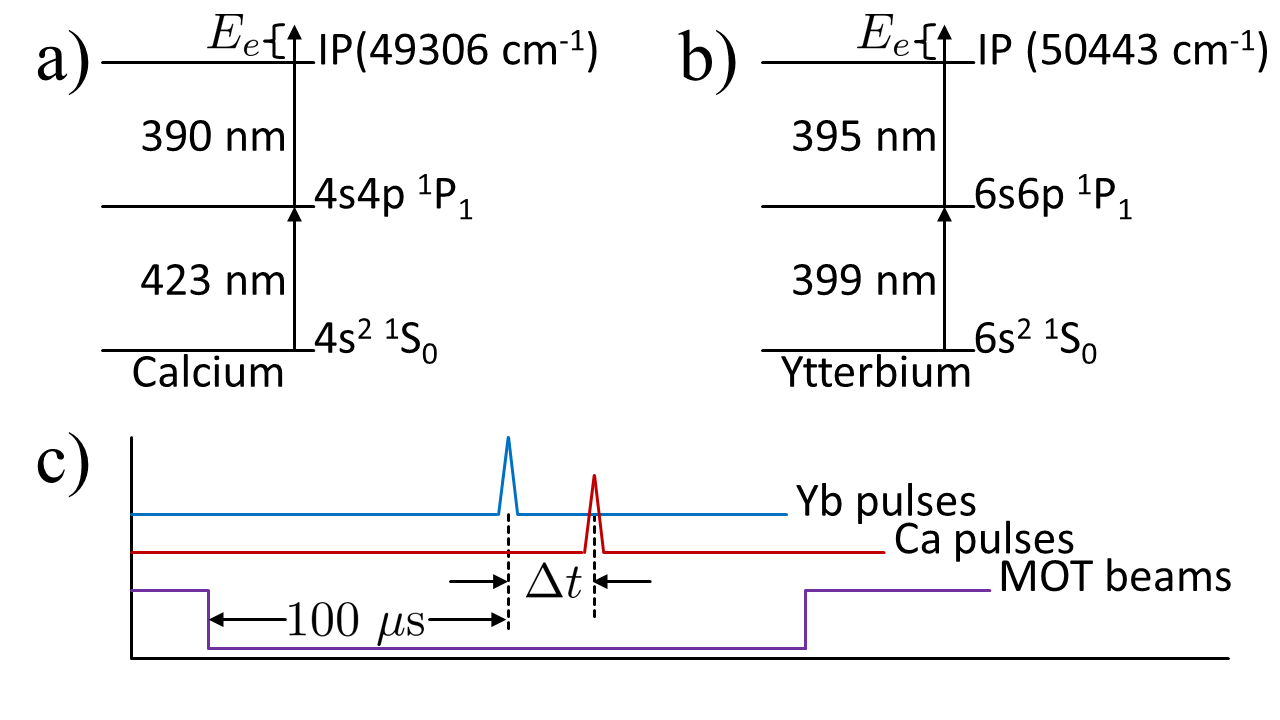}}
  \caption{\label{fig:timingLevels} Partial energy level diagram for Ca (a) and Yb(b) showing the MOT and ionization laser wavelengths. IP = Ionization Potential. $E_e$ = electron energy. A schematic diagram of the experimental timing is also shown (c). The ns-duration laser pulses used to ionize the Yb atoms arrive $\Delta t = 40$~ns before the laser pulses used to ionize the Ca atoms. Energy level information from Ref. \cite{NIST_ASD}. }
\end{figure}

The electron temperature, $T_e$, is determined by the excess photon energy above the atomic ionization potential. Because the electron energy drives the plasma expansion rate, we must characterize $T_e$ accurately. A partial energy level diagram for Ca and Yb is shown in Fig. \ref{fig:timingLevels}. We adjust the wavelength of the 390 nm laser ($\lambda^{-1} = 23754~\mbox{cm}^{-1}$) so that it ionizes Ca atoms out of the $4s4p~^1P_1^{ }$ level ($E_{4s4p}^{\rm Ca} = 23654~\mbox{cm}^{-1}$), imparting $100~\mbox{cm}^{-1}$ of kinetic energy to the electrons. However, this same laser, if it is coincident with the 399 nm Yb excitation laser, would ionize Yb atoms out of the $6s6p~^1P_1^{ }$ level ($E_{6s6p}^{\rm Yb} = 25068~\mbox{cm}^{-1}$), imparting $279~\mbox{cm}^{-1}$ of kinetic energy to the electrons. We minimize this problem in two ways. First, we use fast optical modulators to turn off all of the MOT laser beams 100~$\mu$s before ionizating the MOT. This ensures that none of the Ca and Yb atoms are in the $^1P_1^{ }$ states when the ionizing laser pulses arrive at the MOT. Second, we delay the Ca ionization pulses by 40 ns relative to those for Yb. Because of the larger Yb  mass, this delay is short enough that the Yb$^+$ plasma does not expand before the dual-species plasma is formed. We verify that this delay is equal to 5 times the measured pulse width of the laser pulses and that these precautions prevent spurious ionization to a level below our detection sensitivity.

The time-evolving rms width of the Ca$^+$ ion velocity distribution is determined using laser-induced fluorescence measurements at 397 nm \cite{NIST_ASD}. A linearly-polarized probe laser beam at this wavelength passes through the plasma and is retro-reflected. The single-beam intensity is $I = 50~\mbox{mW/cm}^2 \approx I_{\rm sat}$. A strong laser beam ($I=2000~\mbox{mW/cm}^2$) at 850 nm is used to minimize optical pumping of the Ca$^+$ ions into dark states. The size of these laser beams is large compared to the size of the plasma. Both laser beams illuminate the entire plasma for the duration of the experiment.

When the frequency of the probe laser is detuned by a frequency $\Delta \omega = 2\pi \times (f-f_0)$ from the atomic resonance frequency, $f_0$, the fluorescence signal is proportional to the number of ions Doppler-shifted into resonance with the laser beam. By repeating fluorescence measurements for a range of different probe laser frequencies, we are able to map out the $v_z$ velocity distribution as a function of time, averaged over the entire plasma.

\section{Two-fluid simulation}

The dual-species UNP environment is highly collisional. The Vlasov equation, which has modeled single-species UNP expansion with high accuracy \cite{PhysRevLett.99.155001} is not appropriate. When the low-mass Ca$^+$ ions expand in the presence of the heavier Yb$^+$ plasma, ion friction transfers momentum from calcium to ytterbium, dramatically changing the behavior of the plasma.

To interpret the experimental results, we have built a two-fluid 1-d code in spherical coordinates. A kinetic treatment is almost certainly required, but such a calculation is difficlt and the mean free path in the bulk of the plasma is small enough that a fluid treatment should give some insight. The two species are denoted by the subscript $s$ and it is assumed that the important physical effects are convection, adiabatic expansion, pressure acceleration, acceleration by an ambipolar electric field, and interspecies friction, including Joule heating due to the relative velocity between the two species. Our plasma is not very strongly coupled, so we assume that the monatomic ideal gas law is the equation of state for both species, so that each one has a distribution function approximated by a drifing Maxwellian and so that each species has adiabatic exponent $\gamma=5/3$. For our conditions viscous effects, the thermal force, and ion thermal conduction are small and are not included in the fluid equations. With these assumptions the three equations to be solved for each species are
\begin{align}
\frac{\partial n_s}{\partial t} + u_s \frac{\partial n_s}{\partial r} & = -n_s \nabla \cdot (u_s \hat{r})
 \\
\frac{\partial T_s}{\partial t} + u_s \frac{\partial T_s}{\partial r} & = -\frac{2}{3} T_s \nabla \cdot (u_s \hat{r}) + \frac{2}{3 n_s k_B} Q_{ss^{\prime}} \label{eqn:2fluid}
\\
\frac{\partial u_s}{\partial t} + u_s \frac{\partial u_s}{\partial r} & = -
\frac{k_B}{n_s m_s} \frac{\partial n_s T_s}{\partial r} -
\frac{k_B T_e}{n_s m_s} \frac{\partial n_s}{\partial r} +
\frac{ F_{ss^{\prime}}}{m_s} \label{eqn:3fluid}
\end{align}

In these equations $(n_s,T_s,u_s)$ are, respectively, the density, temperature, and radial fluid velocity for species $s$. The quantity $F_{ss^{\prime}}$ is the interspecies friction force and $Q_{ss^{\prime}}$ is a term representing frictional heating and temperature equilibration between the two species.

To compute the interspecies friction force $F_{ss^{\prime}}^{ }$ and the heating term $Q_{ss^{\prime}}^{ }$, we follow the treatment of Baalrud and Daligault in Ref. \cite{doi:10.1063/1.4967627}, including the energy exchange density in Eqs (44) through (51) of that reference. Our friction force $F_{ss^{\prime}}$ is given in terms of their fluid friction force density ${\bf R}^{ss'}$ by
\begin{equation}
F_{ss'} = \frac{{\bf R}^{ss'}}{n_s} = - \frac{16}{3} \frac{\sqrt{\pi} e^4 n_{s'}}{ (4 \pi \epsilon_0)^2 m_{ss'} \bar{v}_{ss'}^3} \Xi(\Delta \overline{ V })
(\bf{u_s} - \bf{u_{s'}}) \label{eqn:4fluid}
\end{equation}
where $m_{ss^{\prime}}$ is the reduced mass $m_{ss^{\prime}} = m_s m_{s^{\prime}}/(m_s+m_{s^{\prime}})$ and  where
$\bar{v}_{ss^{\prime}}=(2 k_B T_s/m_s + 2 k_B T_{s^{\prime}}/m_{s^{\prime}} )^{1/2}$.
The quantity $\Delta \bar{V}=|{\bf u}_s-{\bf u}_{s^{\prime}}|/\bar{v}_{ss^{\prime}}^{ }$, where ${\bf u}_s$ and ${\bf u}_{s'}$ are the species fluid velocities. The particle velocities ${\bf v}_s$ and ${\bf v}_{s^{\prime}}$ of the two species are assumed to be distributed according to two Maxwellians flowing relative to each other with relative velocity $\Delta V = {\bf u}_s-{\bf u}_{s^{\prime}}$.

The quantity $\Xi( \Delta \overline{ V })$ is a generalized Coulomb logarithm and is given by
\begin{equation}
\label{GenLambda}
  \Xi(\Delta \overline{V })=\frac{3}{16 } \frac{1}{\Delta \overline{ V}^3} \frac{1}{2}  \int_0^{\infty} d \xi \, \xi^2
\frac{\sigma_{ss^{\prime}}^{(1)}(\xi)}{\sigma_0} {\cal{X}},
\end{equation}
where the function ${\cal{X}}$ is
\begin{align}
    {\cal{X}} = [ & (2 \xi \Delta \overline{V} +1) e^{-(\xi + \Delta \overline{V})^2} + \nonumber \\
    &  ( 2 \xi \Delta \overline{V} -1) e^{-(\xi - \Delta \overline{V})^2} ,
\end{align}
where $\sigma_{ss^{\prime}}^{(1)}(\xi) $ is the usual first momentum transfer cross-section \cite{doi:10.1063/1.4967627}, $\xi$ is the ratio of the
particle velocity $v_s$ to the thermal velocity $v_{Ts}=\sqrt{2 k_B T_s /m_s}$, and where
\begin{equation}
\sigma_0 = \frac{\pi e^4}{(4 \pi \epsilon_0)^2 m_{ss^{\prime}}^2 \bar{v}_{ss^{\prime}}^4 }.
\end{equation}
\noindent
Once this friction force is computed we use it in Eqs. (\ref{eqn:2fluid}) and (\ref{eqn:3fluid}) of the fluid model.

Baalrud and Daligault compute the energy exchange and frictional heating term $Q_{ss^{\prime}}$ similarly.
They find
\begin{displaymath}
Q_{ss^{\prime}}=-\frac{16 \sqrt{\pi} n_s n_{s^{\prime}} e^4 k_B }{(4 \pi \epsilon_0)^2 m_s^2 \bar{v}_{ss^{\prime}}^3   } \tilde{\Xi}(\Delta \overline{V}) (T_s - T_{s^{\prime}})
\end{displaymath}
\begin{equation}
- \frac{v_{Ts}^2 }{ \bar{v}_{ss^{\prime}}^2} \Delta {\bf V} \cdot {\bf R}^{ss'}
\end{equation}
where
\begin{equation}
\tilde{\Xi}(\Delta \overline{V}) = \frac{1}{8 \Delta \overline{V} }
\int_0^{\infty} d \xi \xi^4 \frac{\sigma_{ss^{\prime}}^{(1)}(\xi)}{\sigma_0}
\left[ e^{-(\xi-\Delta \overline{V})^2} - e^{-(\xi+\Delta \overline{V})^2} \right]
\end{equation}
This term may then be used in Eq. (\ref{eqn:2fluid})
of the fluid model.

The code is built on a cell-centered spherical grid with $r_i=(i-\textstyle{\frac{1}{2}}) \Delta r$, $i=1, 2, 3, ...$, with $r$ the spherical radial coordinate and with $\Delta r$ the constant grid spacing. We solve these equations using the method of characteristics.

In what follows $i$ denotes the spatial position on the radial grid and $m$ indicates time step in equal time increments $\tau$.
\begin{align}
n_i^{m+1} & = n^m(r_i -\delta r)e^{- \nabla \cdot (u \hat{r}) \tau}
\\
T_i^{m+1} & = T^m(r_i -\delta r)e^{- (2/3) \nabla \cdot (u \hat{r}) \tau} + \nonumber \\ &
\frac{Q}{(2/3) \nabla \cdot (u_s \hat{r}) }
(1-e^{-(2/3) \nabla \cdot (u_s \hat{r}) \tau})
\\
u_i^{m+1} & = u^m(r_i -\delta r) \nonumber \\
& + \left[ -\frac{k_B}{n_s m_s} \frac{\partial n_s T_s}{\partial r} -
\frac{k_B T_e}{n_s m_s} \frac{\partial n_s}{\partial r} +\frac{ F_{s,drag}}{m_s} \right] \tau
\end{align}

We reach back in time from $r_i$ to $r_i-\delta r$ in order to find the quantity to be convected
forward using the approximate characteristic equation
\begin{equation}
\label{drdt}
\dot{r} \approx u_i + u_i'(r-r_i)
\end{equation}
where $u_i$ is the fluid velocity at radial grid point $i$ and where $u_i'$ is the centered approximation to the radial derivative of the fluid velocity at grid point $i$,
\begin{equation}
u_i' = \frac{u_{i+1}-u_{i-1}}{2 \Delta r}
\end{equation}
Solving Eq. (\ref{drdt}) to find the radius from which the density is convected to $r_i$ at $t^{m+1}$ yields for $\delta r$ in $r_i-\delta r$
\begin{equation}
  \delta r = \frac{u_i}{u_i'} (1-e^{-u_i' \tau}) \approx u_i \tau -
\frac{1}{2} u_i u_i' \tau^2
\end{equation}
If $G$ represents any of the quantities $(n,T,v)$ evaluated at the retarded position $r_i - \delta r$, then
\begin{align}
G(r_i -\delta r) & \approx G_i -\frac{\delta r}{2 \Delta r} (G_{i+1}-G_{i-1}) + \nonumber \\
& \frac{\delta r^2}{2 \Delta r^2} (G_{i+1}-2G_i+G_{i-1})
\end{align}

To handle the non-convective parts of the time advance a simple two step predictor-corrector method is used. In the first step old values of $(n_s,T_s,u_s)$ are used to advance to time level $m+1/2$. In the second step these intermediate values are used to advance $(n_s,T_s,v_s)$ to time level $t^{m+1}$.

\subsection{The Coulomb logarithm and momentum transfer}

We have studied two treatments of momentum transfer. The first uses the usual Coulomb cross section, modified by a suitable generalization of $\ln \Lambda$ \cite{PhysRevLett.101.135001, PhysRevE.79.056403, PhysRevLett.110.235001}. As shown in Ref. \cite{doi:10.1063/1.4967627}, for this case the friction force generalized Coulomb logarithm $\Xi(\Delta \overline{V })$ is given by
\begin{equation}
\Xi(\Delta \overline{V }) = \frac{3 \sqrt{\pi}}{4} \frac{\psi(\Delta \overline{V }^2)}{\Delta \overline{V }^3} \ln \Lambda~~~,
\end{equation}
where
\begin{equation}
\psi(x) = {\rm erf} (\sqrt{x}) - \frac{2}{\sqrt{\pi}}\sqrt{x} e^{-x}~.
\end{equation}

Similarly, the Baalrud-Daligault effective Coulomb logarithm for energy exchange
$\tilde{\Xi}(\Delta \overline{V})$ in the case of Coulomb scattering with a Coulomb logarithm multiplier is given by
\begin{equation}
\tilde{\Xi}(\Delta \overline{V}) = \frac{\sqrt{\pi} }{ 2 \Delta \overline{V}} {\rm erf} (\Delta \overline{V}) \ln \Lambda~.
\end{equation}

For the case of electron-ion temperature relaxation, molecular dynamics simulations \cite{PhysRevLett.101.135001} indicate that a Coulomb logarithm of the form,
\begin{equation}
\ln \Lambda = \ln{(1 + C/g)}
\label{eqn:eff}
\end{equation}
is appropriate, where $C=0.7$, and where $g = (e^2/4\pi\epsilon_0)[1/(\lambda_{De} k_B T_e)]$ is the so-called plasma parameter. Effective potential theory calculations suggest that this might be appropriate for our dual-species plasma as well \cite{Baalrud2014}.

Because we are calculating ion-ion momentum transfer, some caution is in order. In the NRL Plasma Formulary, the plasma parameter is $g = r_{\rm min} / r_{\rm max}$. For ion-ion collisions in flowing Maxwellians,
\begin{equation}
    g =
    \frac{e^2}{4\pi\epsilon_0}
    \left[
      \lambda_D \left( \textstyle{\frac{1}{2}} m_{ss'}  \right)
      \left(
        \bar{v}_{ss'}^2 + \textstyle{\frac{2}{3}}
        \left|
          u_s - u_{s'}
        \right|^2
      \right)
    \right]^{-1}
\end{equation}
where the Debye length $\lambda_D$ includes both the ion contribution and a correction due to ion flow and strong coupling, as given in Ref. \cite{PhysRevE.93.043203},
\begin{equation}
  \frac{1}{\lambda_D^2} = \frac{1}{\lambda_e^2} + \sum_i \frac{1}{\lambda_i^2}
  \left(
    \frac{1}{1 + (u_s - u_{s'})^2 / v_{th,i}^2 + 3 \Gamma_i}
  \right),
\end{equation}
where the summation is over the ion species and where $v_{th,i} = (2 k_B T_i/m_i)^{1/2}$. Consistent with Ref. \cite{McQuillen2015} and many other UNP studies, we take the ion strong coupling parameter to be,
\begin{equation}
  \Gamma_i \equiv \frac{e^2}{4\pi\epsilon_0 a_{ws}} \frac{1}{k_B T_i} = 2.3,
\end{equation}
where $a_{ws} = [3/(4\pi n_0)]^{1/3}$ is the Wigner-Seitz radius. Near the center of the plasma, where the density is the highest, the value of the plasma parameter is $g = 2.6$.

The second form for the momentum transfer cross section uses the Debye-screened Coulomb potential described by Stanton and Murillo in Section III and Appendix C, Eq. (41), of Ref. \cite{PhysRevE.93.043203}. The collision integrals in this reference are expressed as convenient functions of the plasma parameter, $g$, as discussed above. Using the screened Coulomb cross section cited above\cite{PhysRevE.93.043203}, the integral in Eq. (\ref{GenLambda}) was performed numerically and fit to an analytic form for use in the fluid code. In this treatment the energy exchange term is density-weighted, as opposed to velocity-weighted in the Baalrud-Daligault treatment, and the temperature equilibration term was neglected since its effect turned out to be small when comparing the simulation to the experiment.

\subsection{The velocity distribution \texorpdfstring{$\bm f(v_z)$}{f(vz)} }

In order to compare the fluid code directly with the experimental measurements, we calculate the $v_z$ velocity distribution from the simulation results for $u(r)$, $n(r)$, and $v_{th}(r)$. In doing so we assume that the particles of each species are drifting Maxwellians with the parameters given in the previous sentence and we integrate over all 3 dimensions in space and over $v_x$ and $v_y$ in velocity space to obtain the following distribution in $v_z$:
\begin{align}
& f(v_z) \propto \nonumber \\
& ~~ \int_0^{\infty} \frac{n(r)}{u(r)} \left[  {\rm erf} \left( \frac{u(r)-v_z}{ v_{th}}\right) +
{\rm erf} \left( \frac{u(r)+v_z}{ v_{th}}\right) \right] r^2 dr,
\label{eqn:fofvz}
\end{align}
where $v_{th} = (2 k_B T /m)^{1/2}$. We evaluate this integral numerically. Note that normalizing constants have been omitted since the experimental data are not normalized.

\section{Comparison of the simulation to the lab data}

\begin{figure}[t]
    \centerline{\includegraphics[width=0.9\linewidth]{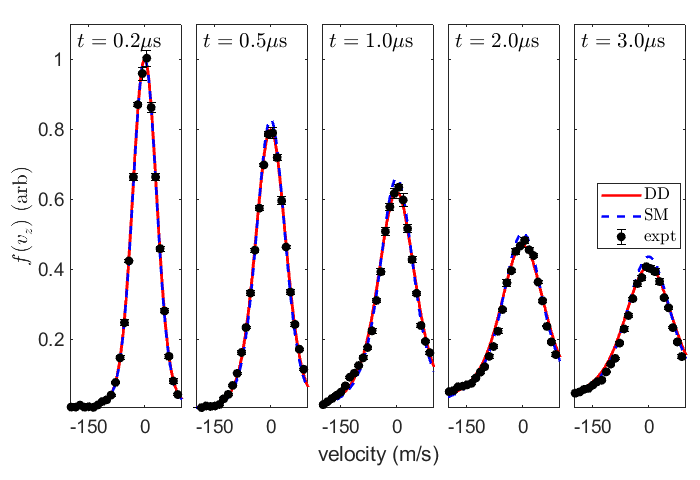}}
    \caption{(Color online) A plot of the velocity distribution $f(v_z^{ })$ at different times. This compares the two-fluid simulations with the experimental measurements of a dual-species Ca/Yb plasma at 0.2, 0.5, 1, 2, and 3 $\mu$s after plasma formation. The Ca and Yb plasmas have the same initial density of $n_0 = 1.8\times 10^{10}~\mbox{cm}^{-3}$. The solid red line uses the Coulomb logarithm of Eq. (\ref{eqn:eff}) with $C=0.7$. The blue dashed line uses the momentum transfer treatment of Ref. \cite{PhysRevE.93.043203}. }
    \label{fig:sm2016}
\end{figure}

\begin{figure}[t]
    \centerline{\includegraphics[width=0.9\linewidth]{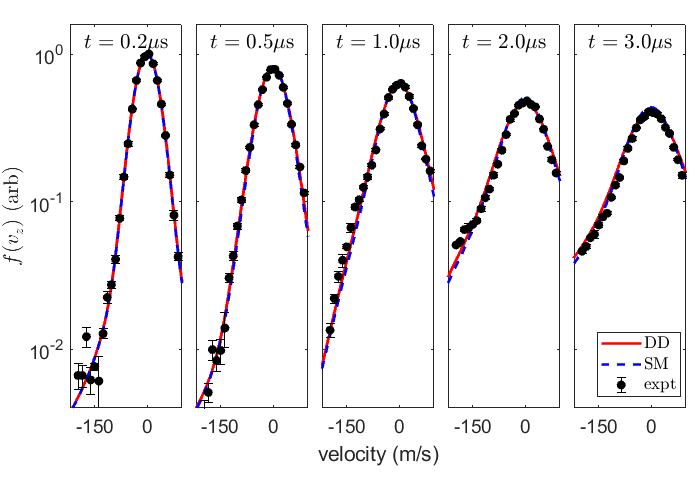}}
    \caption{(Color online) The same data as in Fig. \ref{fig:sm2016} plotted on a semi-logarithmic scale. This representation of the data enables a closer look at the wings of the velocity distribution. The wings of the experimental data rises above the simulations at $1~\mu\mbox{s}$ before falling again at later times.}
    \label{fig:sm20162}
\end{figure}

Before comparing the simulation results to the laboratory measurements we modify the raw data in two ways. First, the simulated velocity distribution is convolved with a Lorentzian line shape. The laboratory data infers the velocity distribution using fluorescence measurements, and those measurements necessarily include a contribution from the natural line width of the atomic transition. The second modification corrects the laboratory measurements for optical pumping. At a given time $t$, the data are multiplied by $\exp(t/\tau)$, where $t$ is the time for which the comparison is made and $\tau$ is estimated to be $30~\mu\mbox{s}$.

In Figs. \ref{fig:sm2016} and \ref{fig:sm20162} we show the simulated and measured $f(v_z)$ distributions at 0.2, 0.5, 1, 2, and 3 $\mu$s after the plasma is formed. The laboratory data are shown in black circles with errorbars indicating the rms noise in the measurements. The red solid line shows the simulated distribution using the effective Coulomb logarithm in Eq. (\ref{eqn:eff}) with $C=0.7$. The blue dashed line shows the distribution from the momentum transfer treatment of Ref. \cite{PhysRevE.93.043203}. As can be seen in Fig. \ref{fig:sm2016}, these two treatments are in good agreement with each other. The uncertainties in the experimental measurements stemming from optical pumping allow both of these simulated distributions to be in agreement with the measurements. We have run the simulation for a range of $C$ values in Eq. (\ref{eqn:eff}). As the value of $C$ increases, the velocity distributions at late times become narrower and the relative values of $f(v_z=0)$ fall less quickly. When $C=1.1$, the velocity distributions generated using the two simulations appear to be in perfect agreement.

A comparison of the data on a logarithmic scale, shown in Fig. \ref{fig:sm20162}, gives a better view of the wings of the distributions. At 0.2 $\mu$s the simulated and measured distributions agree. After 1.0 $\mu$s, the wings of the measured distribution clearly rise above that of the simulations. At later times, that difference becomes less pronounced for the range of velocities that can be measured at present. This difference in the wings is most likely due to kinetic effects not included in the fluid simulation. This is expected, since the mean free path in the edge of the plasma is an appreciable fraction (about 1/3) of the plasma radius. In the wings of the spatial density distribution, the lighter Ca$^+$ ions would be accelerated quickly outwards by the persistent density gradient of the heavier Yb$^+$ ions. Because of the lower density, the friction force would be small. This hypothesis could be tested in the lab using spatial imaging techniques, and that experiment is currently underway.

From the simulation, we can extract information that is not experimentally accessible. During the $3~\mu\mbox{s}$ expansion, the central Yb and Ca densities falls from the initial value of $1.8 \times 10^{10}~\mbox{cm}^{-3}$ to $0.8 \times 10^{10}~\mbox{cm}^{-3}$ for Yb and $0.6 \times 10^{10}~\mbox{cm}^{-3}$ for Ca. The electron temperature, which is assumed to be spatially uniform, falls from 96 K to 45 K.

\begin{figure}[t]
    \centering
    \includegraphics[width=0.9\linewidth]{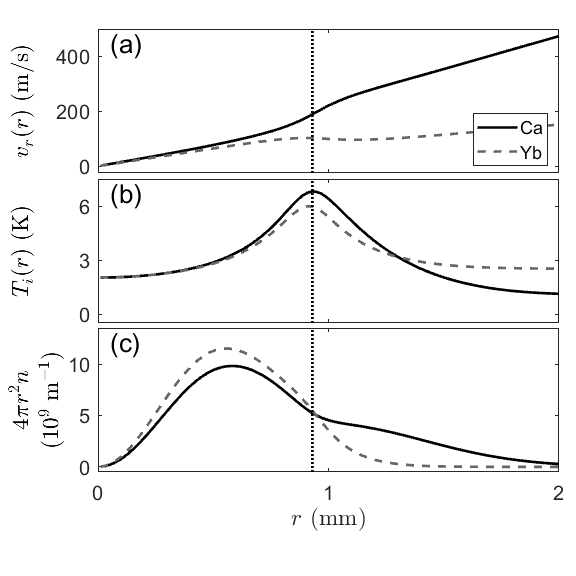}
    \caption{Simulated data showing the radial velocity (a), the ion temperature (b), and the $r$-weighted density (c) after $3.1~\mu$s of plasma expansion using the effective Coulomb logarithm in Eq. (\ref{eqn:eff}) with $C=0.7$. The Ca data is plotted as a solid black line. The Yb data is plotted using a dashed gray line. The vertical dotted line at 0.93 mm is a guide to the eye. The divergence in the relative velocity, the increased ion temperature, and the feature in the $r$-weighted density distribution appear at approximately the same location in space.  }
    \label{fig:vrel}
\end{figure}

The ion radial velocity, density, and temperature depend on the radial coordinate $r$. In the simulation, all three of these parameters develop features near the outside edge of the density distribution. In Fig. \ref{fig:vrel}(a) we show the (radial) flow velocity of the Ca$^+$ and Yb$^+$ ions as a function of $r$ after $3.1~\mu\mbox{s}$. Near the center of the plasma, where the densities are high, the flow velocities match. As the density falls off, the flow velocities diverge.

At the location of the flow-velocity divergence, we also see changes in the ion temperature and density. These quantities are plotted in Fig. \ref{fig:vrel}(b) and (c). The limitations in the simulations suggests these data should not be taken literally. However, these data suggest a cold interior surrounded by a warm shell that is itself surrounded by a cold exterior. We are in the process of setting up an experiment to measure this directly. If these predictions are confirmed, it may be possible to determine $\Gamma$-dependent transport properties in this binary plasma mixture.

\section{Discussion}
We present measurements and simulations of the Ca$^+$ ion velocity distribution in an dual-species UNP of Ca$^+$ and Yb$^+$ ions. The simulation uses two treatments for momentum transfer. One is based on the effective potential approach using the potential of mean force. It uses a Coulomb logarithm extracted from MD simulations. The other uses the momentum transfer treatment of Ref. \cite{PhysRevE.93.043203}, based on the interaction of Yukawa-screened charges. These momentum transfer treatments are included in the friction force between two flowing Maxwellian distributions of Ca$^+$ and Yb$^+$ ions at the same temperature.

Each of these treatments result from different assumptions. The effective potential approach using the potential of mean force is most appropriate for near-equilibrium processes. It assumes that the ion pair distribution function can be calculated using thermodynamic considerations. After the initial disorder-induced heating process in the UNP, this assumption is almost certainly valid. In the present work, we have used a Coulomb logarithm derived from MD simulations based on the effective potential approach. Because that work studied electron-ion energy relaxation, that Coulomb logarithm should be appropriate for momentum transfer processes. However, because we are studying momentum transfer between ions of different mass and not between electrons and ions, we have modified the parameters used in the Debye length and in the plasma parameter, as described previously. Future work should use the full theory in order to remove the approximations used in the present study.

The treatment of Ref. \cite{PhysRevE.93.043203} assumes that the ion-ion potential is accurately represented using a Yukawa-screened interaction. For small values of $\Gamma$ (or large values of $g$) when the plasma is not strongly-coupled, this approach works nicely. As the plasma becomes strongly-coupled, the screening length is modified and Yukawa screening is assumed to be largely correct.

It might be useful to compare these approaches with laboratory data in which the coupling parameter $\Gamma$ is larger, where the underlying assumptions in the theoretical approaches could be tested more directly. For example, one can imagine an experiment in which the Ca plasma is generated and allowed to expand. As it expands, the ion-ion coupling parameter increases to values near 5 \cite{McQuillen2015}. The co-located Yb plasma could then be generated and the interaction between the cold Ca$^+$ ions and the hot Yb$^+$ ions could be measured. Alternatively, very recent work demonstrated the successful laser-cooling of ions in a strontium UNP \cite{Langin2019}. That method could be used to reach $\Gamma = 11$ in one species while observing collisions and interactions due to the presence of the other, with the perturbing plasma either at the same or at higher temperatures.

On the other hand, experiments at higher $\Gamma$ values might produce only trivial transport results. Strong coupling should result in small values of the transport coefficients due to the greater collisional locking or caging of the plasma ions. Such predictions could be verified in our dual-species ultracold neutral plasma. Either way, the laser cooling or heating in UNPs shown in Ref. \cite{Langin2019} could provide a convenient way to measure $\Gamma$-dependent transport in a tightly-controlled environment.

The simulations show that the Ca$^+$ flow velocity matches the Yb$^+$ flow velocity in the center of the plasma. One can envision an experiment in which that flow could be interrupted, or in which two somewhat spatially-offset plasmas could approach equilibrium, flowing over each other. Perhaps flow-related instabilities could be observed. The simulations also show ion heating in regions where the spatial density gradient increases. Experiments are underway now to study this effect.

\section{Conclusion}

In conclusion, we report the first experimental realization of a two-ion species  ultracold neutral plasma, comprised of Yb$^+$ and Ca$^+$ ions and electrons. The mm-sized spherical plasma is not confined but expands radially under the influence of the ambipolar field. We measure the spatially-averaged velocity distribution of the Ca$^+$ ions as the plasma expands and observe that the Yb$^+$ ions significantly slow the rate at which the distribution broadens. This results from momentum transfer between the Yb$^+$ and Ca$^+$ ions in the plasma.

We compare these measurements with the output of a fluid-code simulation. In the simulation we use two different expressions for the momentum transfer cross section. One derives from the potential of mean force. The other derives from a coupling-corrected Yukawa interaction. In the fluid equations, momentum transfer manifests most strongly as a friction force between flowing (assumed) Maxwellian distributions of Ca$^+$ and Yb$^+$ ions at the same temperature. The momentum transfer cross section traditionally includes a Coulomb logarithm, that, in our system, has a value less than 1.

We find that both formulations of the momentum transfer cross section, when included in our fluid simulation, produce nearly identical radial velocity distributions. The main features of the simulated distribution match the measured velocity distribution well. Some behavior in the wings of the distribution are noted, perhaps due to kinetic effects outside of the fluid code assumptions. It is possible that differences between these two momentum transfer treatments might  appear if spatially-resolved measurements could be made in plasmas of varying levels of strong Coulomb coupling. It is  possible that higher fidelity simulations might also reveal differences. Experiments and calculations are currently underway to test these ideas.

\section{Acknowledgements}
This project was supported by grants from the National Science Foundation (Grant No. PHY-1500376) and the Air Force Office of Scientific Research (AFOSR FA9550-17-1-0302). The authors express appreciation to Scott Baalrud and Jerome Daligault for helpful discussions concerning the effective potential approach.

\end{document}